\documentstyle[12pt,epsf]{article}
\renewcommand{\baselinestretch}{1.0}

\newcommand{\be}{\begin{equation}}
\newcommand{\ee}{\end{equation}}
\begin{document}
\topmargin 0pt
\oddsidemargin=-0.4truecm
\evensidemargin=-0.4truecm
\renewcommand{\thefootnote}{\fnsymbol{footnote}}

\newpage
\setcounter{page}{1}
\begin{titlepage}
\vspace*{-2.0cm}
\begin{flushright}
\vspace*{-0.2cm}

\end{flushright}
\vspace*{0.5cm}

\begin{center}
{\Large \bf Homestake result, Sterile neutrinos and Low energy solar
neutrino experiments}
\vspace{0.5cm}

{P. C. de Holanda$^{1}$ and  A. Yu. Smirnov$^{2,3}$\\

\vspace*{0.2cm}
{\em (1) Instituto de F\'\i sica Gleb Wataghin - UNICAMP,
13083-970 Campinas SP, Brazil}\\
{\em (2) The Abdus Salam International Centre for Theoretical Physics,
I-34100 Trieste, Italy }\\
{\em (3) Institute for Nuclear Research of Russian Academy
of Sciences, Moscow 117312, Russia}

}
\end{center}

\vskip 1cm

\begin{abstract}
The Homestake result is about $\sim 2 \sigma$ lower than 
the $Ar$-production rate, $Q_{Ar}$, predicted by  
the LMA MSW solution of the solar neutrino problem.
Also there is no apparent upturn of the energy spectrum 
($R \equiv N_{obs}/N_{SSM}$) at low energies in SNO and Super-Kamiokande.  
Both these facts can
be explained if a light, $\Delta m^2_{01} \sim (0.2 - 2) \cdot 10^{-5}$
eV$^2$, sterile neutrino exists which mixes very weakly with active
neutrinos: $\sin^2 2\alpha \sim (10^{-5} - 10^{-3})$.  We perform both
the analytical and numerical study of the conversion effects in the system
of two active neutrinos with the LMA parameters and one weakly mixed
sterile neutrino. The presence of sterile neutrino leads to a dip in
the survival probability in the intermediate energy range $E = (0.5 -
5)$ MeV thus suppressing the $Be$, or/and $pep$, $CNO$ as well as $B$ electron 
neutrino fluxes.  Apart from diminishing $Q_{Ar}$ it leads to
decrease of the $Ge$-production rate and may lead to decrease of the
BOREXINO signal and CC/NC ratio at SNO.  Future studies of the solar
neutrinos by SNO, SK, BOREXINO and KamLAND as well as by the new low energy
experiments will allow us to check this possibility.  We present a
general analysis of modifications of the LMA energy profile
due to mixing with new  neutrino states. 

\end{abstract}

\end{titlepage}
\renewcommand{\thefootnote}{\arabic{footnote}}
\setcounter{footnote}{0}
\renewcommand{\baselinestretch}{0.9}

\section{Introduction}

In the assumption of CPT invariance  the first KamLAND result~\cite{kam} 
and the results of SNO salt phase~\cite{salt} 
confirm the large mixing angle (LMA) MSW solution of the solar neutrino problem
\cite{w1,ms,lma}.  Is the LMA solution complete? If there are observations which
may indicate some deviation from LMA?

According to the recent analysis, LMA MSW describes all the data very
well~\cite{solar,comb}: pulls of predictions from experimental results
are below $1\sigma$ for all but the Homestake experiment~\cite{comb}.  The  generic
prediction of LMA for the $Ar$ production rate is
\be
Q_{Ar} = 2.9 - 3.1~~ {\rm SNU}, 
\ee
which is about $2\sigma$ higher than the Homestake result~\cite{hom}.  This pull can be

\begin{itemize}

\item
just a statistical fluctuation;

\item
some systematics which may be related to the claimed long term time
variations of the Homestake signal~\cite{hom}; 

\item
a consequence of higher fluxes predicted by the Standard Solar Model
(SSM) \cite{ssm}~\footnote{For instance, the $CNO$-neutrino fluxes
have rather large uncertainties. According to the SSM and in the LMA
context they contribute  to  $Q_{Ar}$ about 0.25 SNU, so that reduction of the
$CNO$- fluxes by factor of 2 (which is within $2\sigma$ of the
estimated uncertainties) leads to reduction of the $Ar$-production
rate by $\Delta Q_{Ar} \sim 0.12$ SNU.},
 
\item 
some physics beyond LMA.

\end{itemize}

Another generic prediction of LMA is the ``upturn" of the energy spectrum 
at low energies 
(the upturn of ratio of the observed and the SSM predicted numbers of events).  
According to LMA, the survival probability should increase
with decrease of energy below (6 - 8) MeV ~\cite{lma}.  For the best
fit point the upturn can be as large as 10 - 15  \% between 8 and 5
MeV~\cite{ks,comb}.  Neither Super-Kamiokande (SK)~\cite{sk} nor
SNO~\cite{sno} show the upturn, though the present sensitivity is not
enough to make statistically significant statement.

There are also  claims that the solar neutrino data have 
time variations with small periods ~\cite{vari}. If true, this 
can not be explained in the context of LMA solution.

Are these observations related? Do they indicate some new physics in
the low energy part of the solar neutrino spectrum? 
In this paper we show that both the lower $Ar$-production rate and the
absence of (or weaker) upturn of the spectrum can be explained by the
effect of new (sterile) neutrino.  The solar neutrino conversion 
in the non-trivial $3\nu$- context (when the effect of  third
neutrino is not reduced to the averaged oscillations) have been
considered in a number of publications before~\cite{ms,three}. In
particular, modification of the $\nu_e$- survival probability 
due to  the mixing with sterile neutrino has been studied~\cite{3st}.  Here we suggest
specific parameters of the sterile neutrino which lead to appearance of a
dip in the adiabatic edge of the survival probability ``bath", 
at $E = (0.5 - 2)$ MeV, and/or  flattening of the spectrum distortion 
at higher energies $(2 - 8)$ MeV.  The dip
suppresses the $Be$- $(\nu_e)$ neutrino flux or/and other fluxes at
the intermediate energies, and consequently, diminishes the
$Ar$-production rate. It also diminishes or eliminates completely
(depending on mixing angle and $\Delta m^2_{01}$) the upturn of
spectrum.  We comment on a possibility to induce time variations of
signals by the presence of very small mixing with sterile neutrinos.

The paper is organized as follows. In the Sec.~2 we introduce mixing
with sterile neutrino and  study in sec.~3 
both analytically and numerically the conversion as well as  the energy profile of the effect.
In Sec.~4 physical consequences of the modification of the
energy profile are considered.  We calculate predictions for observables,  the
$Ar$-production rate, the $Ge$-production rate, the CC/NC ratio at SNO
and the rate at BOREXINO, as functions of the mixing and mass of
sterile neutrino in sec.~5.  We consider an impact of the sterile neutrino on
the global fit of the solar neutrino data in sec. 6, where we describe  three 
possible scenario in sec.~6. 
In sec.~7 we discuss
future checks of the suggested scenarios.  We present a general
analysis of possible modifications of the LMA profile by mixing with
additional neutrino states in the Appendix.  Our results
are summarized in sec.~8.

\section{Sterile neutrino mixing and   level crossing}

Let us  consider the  system of two active neutrinos, $\nu_e$ and $\nu_a$, and one
sterile neutrino, $\nu_s$, 
which mix in the mass eigenstates $\nu_1$,
$\nu_2$ and $\nu_0$:
\begin{eqnarray}
\nu_0 &=& \cos \alpha~\nu_s + \sin \alpha (\cos\theta~\nu_e  - \sin\theta~\nu_a), \nonumber\\
\nu_1 &=&  \cos \alpha~ (\cos\theta~\nu_e  - \sin\theta~\nu_a) - \sin \alpha~\nu_s, 
\nonumber\\
\nu_2 &=& \sin\theta~\nu_e + \cos\theta~\nu_a~.
\label{mix}
\end{eqnarray}
The states $\nu_e$ and $\nu_a$ are characterized by the LMA
oscillation parameters, $\theta$ and $\Delta m^2_{12}$. They mix in
the mass eigenstates $\nu_1$ and $\nu_2$ with the eigenvalues $m_1$, and $m_2$.
The sterile neutrino is mainly present in the mass eigenstate $\nu_0$
(mass $m_0$). It mixes weakly ($\sin \alpha \ll 1$) with active
neutrinos in the mass eigenstate $\nu_1$
\footnote{The introduction of mixing with the third active neutrino is straightforward. 
This mixing (if not zero) can produce small averaged oscillation
effect and in what follows it will be neglected.}.  We will assume
first that $m_2 > m_0 > m_1$ and consider  the oscillation parameters of
$\nu_s$ in the intervals:
\be
\Delta m^2_{01} = m^2_0 - m_{1}^2  = (0.2 - 2) \times 10^{-5} ~~{\rm  eV^2},~~~~  
\sin^2 2\alpha \sim  10^{-5} - 3 \cdot 10^{-3}.
\ee

Let $\nu_{1m}$, $\nu_{2m}$, $\nu_{0m}$ be the eigenstates,
and $\lambda_1$, $\lambda_2$, $\lambda_0$ the corresponding
eigenvalues of the $3\nu$-system in matter.  We denote  the
ratio of mass squared differences as 
\be
R_{\Delta} \equiv \frac{\Delta m^2_{01}}{\Delta m^2_{21}}.
\ee

\begin{figure}[ht]
\centering\leavevmode
\epsfxsize=.8\hsize
\epsfbox{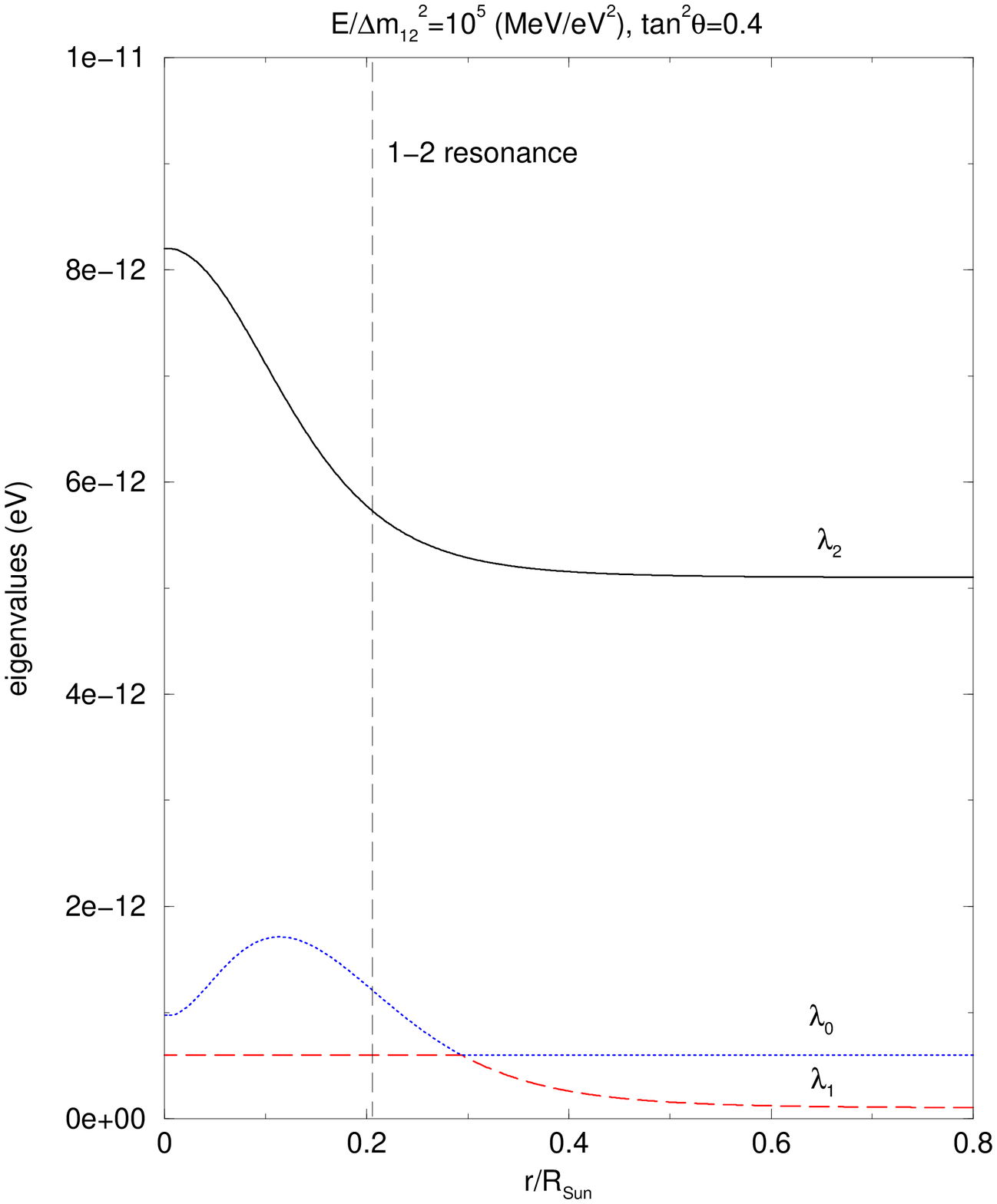}
\caption{The level crossing scheme. 
The  mass eigenvalues as  functions of the distance from the center of the Sun 
  for  $E/\Delta m^2_{12}=10^{5}$ eV$^2$ 
  and $\tan^2\theta=0.4$. The mass ratio is taken to be 
  $R_{\Delta}=\Delta m_{01}^2/\Delta m_{12}^2 = 0.10$. Also shown
is  the position of 1-2 resonance (dashed  vertical line).
}
\label{fig:eigenvalues}
\end{figure}

The level crossing scheme, that is, the dependence of
$\lambda_i$, ($i = 0, 1, 2$) on the distance inside the Sun (or
on the density), is shown in fig.~\ref{fig:eigenvalues}.  It can be
constructed analytically considering mixing of the sterile neutrino
(the $s$-mixing) as a small perturbation.

1). In the  absence of $s$-mixing  we have  usual LMA
system of two active neutrinos with eigenstates $\nu_{1m}^{LMA}$,
$\nu_{2m}^{LMA}$, and the eigenvalues $\lambda_1^{LMA}$ and
$\lambda_2^{LMA}$ which we will call the LMA levels:
\be 
\lambda_1^{LMA} = \frac{m_1^2+m_2^2}{4E} + \frac{V_e + V_a}{2} -
\sqrt{ \left( \frac{\Delta m_{21}^2}{4E}\cos 2\theta - \frac{V_e - V_{a}}{2} \right)^2 +
\left(\frac{\Delta m_{21}^2}{4E}\sin 2\theta \right)^2},  
\label{lambda} 
\ee 
and $\lambda_2^{LMA}$ has similar expression with plus sign in front
of square root. Here $V_e = \sqrt{2} G_F (n_e - 0.5 n_n$), and $V_{a}
= - 0.5 \sqrt{2} G_F n_n$ are the matter potentials for the electron
and non-electron active neutrinos respectively; $n_e$ and $n_n$ are
the number densities of the electrons and neutrons. For the sterile
neutrino we have $V_s =0$. The 1-2 (LMA) resonance condition
determines the LMA resonance energy:
\be
E_a = \frac{\Delta m_{21}^2\cos 2\theta}{2(V_e - V_{a})}.
\ee

2). Let us turn on the $\nu_s$-mixing.  In the assumption $m_1 < m_0 <
m_2$ the sterile neutrino level $\lambda_s$ crosses $\lambda_1^{LMA}$
only.  The level $\lambda_2^{LMA}$ essentially decouples. It is not
affected by the $s$-mixing, and $\lambda_2 \approx \lambda_2^{LMA}$.
Evolution of the corresponding eigenstate, $\nu_{2m}$, is strongly adiabatic.

3). In general, the sterile level, $\lambda_s$, as the function of
density, crosses $\lambda_1^{LMA}$ twice: above and below the 1-2
resonance density.  Effects of the higher (in density) level crossing
can be neglected since the neutrinos of relevant energies are produced
at smaller densities.  This can be seen in the
fig.~\ref{fig:eigenvalues} where the second crossing of
$\lambda_1^{LMA}$ and $\lambda_s$ would be on the left, if the density
would continue to increase above the central solar density.
Consequently, there are two relevant resonances in the system 
associated with  1-2 level crossing  
(the LMA resonance) and  with  1-s crossing.  For
low energies (below the $s$-resonance) $\lambda_1 \approx
\lambda_1^{LMA}$ and $\lambda_0 \approx
\lambda_s$.

The Hamiltonian of the ($\nu_{1m}^{LMA} - \nu_s$)
sub-system can be obtained diagonalizing the $\nu_e - \nu_a$ block of
of the $3\nu$ Hamiltonian, and then neglecting small 1-3 element. As a result
\be 
H = \left(
\begin{array}{ll} 
\lambda_1^{LMA} & \frac{\Delta m_{01}^2}{4E}\sin 2\alpha \cos(\theta - \theta_m)\\ 
\frac{\Delta m_{01}^2}{4E}\sin 2\alpha \cos(\theta - \theta_m) & 
\frac{m_1^2+m_0^2}{4E}+\frac{\Delta m_{01}^2}{4E}\cos 2\alpha  
\end{array} \right), 
\label{ham} 
\ee 
where $\lambda_1^{LMA}$ is given in (\ref{lambda}).  The $1- s$
resonance condition, 
\be
\lambda_1^{LMA}(\Delta m_{21}^2/E,  \theta, V_e, V_a)  = 
\frac{m_1^2+m_0^2}{4E}+\frac{\Delta m_{01}^2 \cos 2\alpha}{4E}, 
\label{res}
\ee 
determines the $s$-resonance energy
$$
E_s = \frac{0.5 m^2_1 + \Delta m_{01}^2 \cos^2\alpha}{V_e + V_a}
\times ~~~~~~~~~~~~~~~~~~~~~~~~~~~~~~~~~~~~~~~~~~~~~~
$$
\be
\times \frac{1 - R_{\Delta}}{1 - 2R_{\Delta} + \xi \cos 2\theta +    
\sqrt{(1 - 2R_{\Delta} + \xi \cos 2\theta )^2 - 4 R_{\Delta}(1 - R_{\Delta}) (\xi^2 -1)}}, 
\label{e-res}
\ee
where $\xi \equiv (V_e - V_a)/(V_e + V_a) = n_e/(n_e - n_n)$.  Notice
that since $\lambda_1^{LMA}$ is a non-linear function of the neutrino
energy,  the proportionality  $E_s \propto \Delta m_{01}^2$ 
is broken and $E_s$ turns out to be complicated function of
$\Delta m_{01}^2$.

Another feature of the $\nu_{1m}^{LMA} - \nu_s$ sub-system is that due to dependence of
$\theta_m$ on $E$, the effective mixing parameter in (\ref{ham}),
$\propto \sin 2\alpha \cos(\theta - \theta_m)$, also depends on the energy
(decreases with $E$), though this dependence is weak.  Indeed, in
the case of small $\alpha$ and the $s$-resonance being substantially below the
LMA resonance,  we can take $\theta \approx \theta_m$ in the first approximation,  
$\cos(\theta - \theta_m) \approx 1$.  Even in the LMA
resonance, when $\theta_m = \pi/2$, we get $\cos(\theta - \theta_m) =
0.97$. \\

\section{Survival probability. Properties of the dip}

Let us find the $\nu_e$ survival probability.  According to
(\ref{mix}) the initial neutrino state can be written in terms of the
matter eigenstates $\nu_{im}$ as
\be   
\nu_e = \sin\theta_m^0 \nu_{2m} + \cos \theta_m^0 
(\cos\alpha_m^0 \nu_{1m} +
\sin \alpha_m^0 \nu_{0m}), 
\label{nuem}
\ee
where $\theta_m^0$ and $\alpha_m^0$ are the mixing angles in matter in
the neutrino production point.  

Propagation of neutrinos from the production point
to the surface of the Sun is described in the following way. 
$\nu_{2m}$ evolves adiabatically, so that $\nu_{2m} \rightarrow
\nu_{2}$.  Evolution of the two other eigenstates is, in general,
non-adiabatic, so that
\be
\nu_{1m} \rightarrow A_{11} \nu_1 + A_{01} \nu_0,~~~~~~\nu_{0m} \rightarrow A_{10} \nu_1 + A_{00} \nu_0, 
\label{0tra}
\ee
where $A_{ij}$ are the transition amplitudes which satisfy the
following equalities: $|A_{01}|^2 = |A_{10}|^2 = 1 - |A_{00}|^2 = 1 -
|A_{11}|^2 \equiv P_2$. They can be found by solving the evolution
equation with the Hamiltonian (\ref{ham}).  $P_2$ is the two neutrino
jump probability in the system $\nu_{1m} - \nu_s$.

Using (\ref{nuem}, \ref{0tra}) we can write the final neutrino state
as
\be   
\nu_f = \sin\theta_m^0 \nu_{2} e^{i\phi_2} + 
\cos \theta_m^0 \left[
\cos\alpha_m^0 (A_{11} \nu_1 + A_{01} \nu_0) +           
\sin \alpha_m^0 (A_{10} \nu_1 + A_{00} \nu_0) \right], 
\label{nuf}
\ee
where $\phi_2$ is the phase acquired by $\nu_{2m}$.  Then the survival
probability is given by
\be
P_{ee} \equiv |\langle \nu_e |\nu_f \rangle|^2 \approx \sin^2
\theta_m^0 \sin^2 \theta +
\cos^2 \theta_m^0 \cos^2 \theta \left[\cos^2 \alpha_m^0 - P_2 \cos 2\alpha_m^0\right].  
\label{prob}
\ee
Here we have neglected a small admixture of $\nu_e$ in $\nu_0$: $\langle
\nu_e |\nu_0 \rangle \approx 0$. Also we have taken into
account that the coherence of the mass eigenstates is destroyed on the
way from the Sun to the Earth due to a spread of the wave packets and
averaging effects.

Similarly we obtain the transition probability of the electron to sterile neutrino: 
\be
P_{es} \equiv |\langle \nu_s |\nu_f \rangle|^2 \approx
\cos^2 \theta_m^0 \left[\sin^2 \alpha_m^0 +  P_2 \cos 2\alpha_m^0\right].
\label{probs}
\ee

Let us consider specific limits of the formula (\ref{prob}).  If  
evolution is adiabatic in the $s$-resonance (which can be realized
for the large enough $s$-mixing), we find  $P_2 = 0$ and
\be   
P_{ee} = \sin^2 \theta_m^0 \sin^2\theta +
\cos^2\theta_m^0 \cos^2\theta \cos^2\alpha^0_m .          
\label{nuead}
\ee
In the opposite case of strongly non-adiabatic conversion ($P_2 \approx 1$) the probability equals 
\be   
P_{ee} \approx \sin^2\theta_m^0 \sin^2\theta +
\cos^2\theta_m^0 \cos^2\theta \sin^2\alpha^0_m .           
\label{nuenad}
\ee
Notice that in spite of strong violation of adiabaticity 
in the $s$-resonance, the effect of $s$-mixing is still
present due to the averaging of oscillations.

The energy dependences of the probabilities can be easily understood
using the results given in Eqs. (\ref{prob} - \ref{nuenad}).  Let $E_a
(n_c)$ and $E_s(n_c)$ be the LMA resonance energy and the $s$-resonance
energy which correspond to the central density of the Sun $n_c$.
Then the following consideration holds. 

1). For high energies, $E > E_a(n_c)$, 
neutrinos are produced far above the 1-2 resonance density, so that
$\theta_m^0 \approx \pi/2$.  Then according to (\ref{prob}), $P =
\sin^2 \theta$, as in the $2\nu$ case, independently of properties of
the $s$-resonance.  The initial state coincides practically with
$\nu_{2m}$, and the later  propagates adiabatically.

The $s$-resonance becomes operative at the energies of adiabatic edge, when
$\theta_m^0$ deviates from $\pi/2$. This is the consequence of the
fact that $\lambda_s$ crosses the lowest LMA level $\lambda_1$. 

2).  For low energies, $E < E_s(n_c)$, the $s$-resonance is not
realized inside the Sun and $s$-mixing equals the
vacuum mixing ($\cos^2 \alpha^0_m \approx \cos^2 \alpha^0 \approx 1$).
Then from (\ref{prob}) we get the usual adiabatic formula for the $2\nu$
case
\be   
P_{ee}^{adiab} \approx \sin^2 \theta_m^0 \sin^2 \theta +
\cos^2 \theta_m^0 \cos^2 \theta~.            
\label{esmall}
\ee

3). At the intermediate energies, crossing 
the $s$-resonance can be adiabatic (at $E \sim E_s(n_c)$), and moreover,
the initial angle can be equal to $\alpha^0_m \approx \pi/2$. Since
the $s$-resonance is very narrow this equality is realized already at
energies slightly above $E_s(n_c)$. In this case we get from
(\ref{nuead})
\be   
P_{ee} \approx \sin^2 \theta_m^0 \sin^2 \theta.
\label{nearmin}
\ee
If also $E \ll E_a(n_c)$, so that $\theta_m^0 \approx \theta$, the
Eq.~(\ref{nearmin}) leads to
\be   
P_{min} = P_{ee} \approx \sin^4 \theta~.
\label{pmin}
\ee
$P_{min}$ is the absolute minimum of the survival probability which can be
achieved in the system. In general, $P_{ee} > \sin^4 \theta$,  since $E$
is not small in comparison with $E_a$ ($\sin \theta_m^0 > \sin
\theta$) and/or the adiabaticity is broken.

For $\alpha^0_m \approx \pi/2$, which can be realized  for $E$ 
being slightly higher than $E_s$, we find from (\ref{prob})
\be   
P_{ee} \approx \sin^2 \theta_m^0 \sin^2 \theta +
\cos^2 \theta_m^0 \cos^2 \theta \sin^2 \alpha^0_m P_2 .         
\label{nuead2}
\ee
With the increase of energy the adiabaticity is violated, $P_2
\rightarrow 1$, and the probability approaches the adiabatic one for
the $2\nu$ system (\ref{esmall}). \\

\begin{figure}[ht]
\centering\leavevmode
\epsfxsize=.7\hsize
\epsfbox{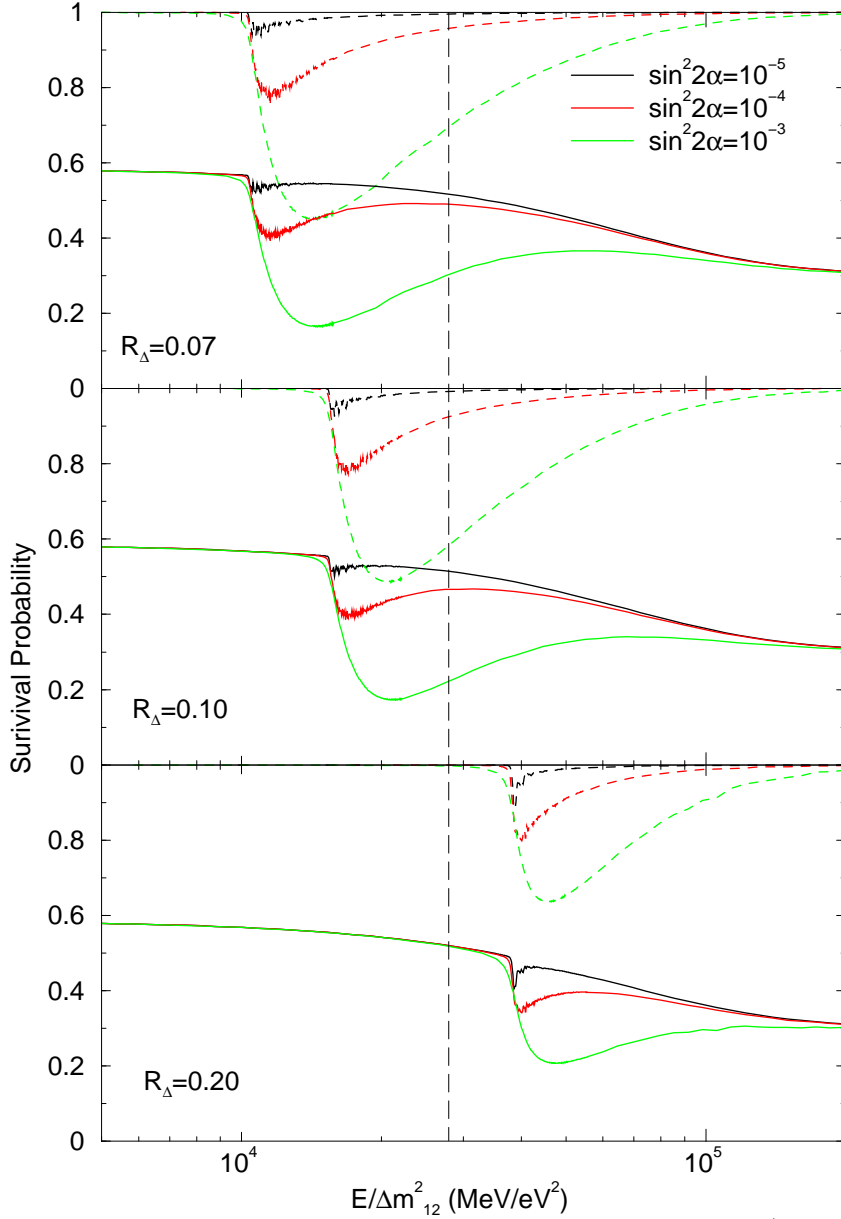}
\caption{
The survival probability of the electron neutrinos,  
$P_{ee}$,  (solid line) and survival probability 
of the active neutrinos, $1 - P_{es}$, (dashed line), 
as functions of $E/\Delta m_{12}^2$  
for different values of the sterile-active mixing  
parameter $\sin^2 2\alpha$. We take $\tan^2\theta = 0.4$. 
Also shown is  position of the 1-2 resonance  for the central density of the Sun.
(vertical dashed line). For $\Delta m_{12}^2 = 7.1 \cdot 10^{-5}$ eV$^2$ the $Be$-line is at 
$E/\Delta m_{12}^2 = 1.2 \cdot 10^{4}$ MeV/eV$^2$,  the $pep$- neutrino line
is at $E/\Delta m_{12}^2 = 2\cdot 10^{4}$ MeV/eV$^2$,  the lowest
(observable) energy, $E = 5$ MeV, and the highest energy of boron
neutrino spectrum ($\sim 14$ MeV) are at $E/\Delta m_{12}^2 = 7 \cdot
10^{4} $ MeV/eV$^2$ and $2 \cdot 10^{5} $ MeV/eV$^2$
correspondingly.  
}
\label{fig:psurv}
\end{figure}

In fig.~\ref{fig:psurv} we show results of numerical
computations of the $\nu_e$ survival probability $P_{ee}$,  and the
survival probability of active neutrinos, $(1 - P_{es})$,  as
functions of energy.  In our numerical calculations we have performed
a complete integration of the evolution equations for the $3\nu$-system
and also made  averaging over the production region of
the Sun.  The analytical consideration allows us to understand
immediately the numerical results shown in fig. \ref{fig:psurv}.

The effect of $s$-mixing is reduced to  appearance of a  dip in the
LMA energy profile.  A size of the dip equals:
\be   
\Delta P_{ee} \equiv P_{ee}^{LMA} - P_{ee} =   P_{es} \cos^2\theta,    
\label{delta}
\ee
where $P_{ee}(E)^{LMA} = P_{ee}(E)^{adiab}$ is the LMA probability
given by the adiabatic formula (\ref{prob}).  To obtain the last equality in
(\ref{delta}) we used  expressions for $P_{ee}$ from (\ref{prob}),
$P_{ee}(E)^{LMA}$ - from (\ref{esmall}) and $P_{es}$ - from (\ref{probs}). 
Since $\cos^2\theta < 1$  
(the best fit value of LMA mixing, $\cos^2\theta = 0.714$)   
according to (\ref{delta}) a change of the $\nu_e$
survival probability due to mixing with $\nu_s$ is weaker than
the transition to sterile neutrino $P_{es}$.  
The relation (\ref{delta}) is well reproduced in fig.~\ref{fig:psurv}.

A position of the dip (its low energy edge) is given by the
resonance energy taken at the central density of the Sun $E_s (n_c)$ (\ref{e-res}). 
With increase of $\Delta m_{01}^2$ the dip shifts
to higher energies. However, this shift is stronger than simple
proportionality to $\Delta m_{01}^2$ as can be found from  (\ref{e-res}).  For
instance, the increase of $R_{\Delta}$ from 0.1 to 0.2 leads to the
shift of dip by factor 2.6 in the energy scale (see
fig.~\ref{fig:psurv}).

The maximal suppression in the dip depends on $R_{\Delta}$ and 
$\alpha$. For small $R_{\Delta}$ (large spit between the two
resonances) and large $\alpha$ ($\sin^2 2\alpha > 10^{-3}$) the
absolute minimum (\ref{pmin}) can be achieved. Indeed, the condition
for the minimum is nearly satisfied for the solid line  in the upper panel
of fig.~\ref{fig:psurv} where $P_{ee} \sim 0.1$.

With increase of $R_{\Delta}$ (smaller split of the resonances) or/and
decrease of  $\alpha$ (stronger violation of the adiabaticity) a 
suppression in the dip weakens.  Also with decrease of $\alpha$ 
the dip becomes narrower.

Similarly one can consider crossing of $\lambda_s$ with 
$\lambda_2^{LMA}$. In this case the effect on $P_{ee}$ is weaker due
to smaller admixture of $\nu_e$ in $\nu_2$.  Now the dip can
appear at higher energies in the non-oscillatory part of the LMA profile
where $P_{ee} \approx \sin^2\theta$.

\section{Observables and restrictions}

As follows from fig.~\ref{fig:psurv}, selecting appropriately  the values of  
$R_{\Delta}$ and $\alpha$ (and therefore  position and form of the dip)
one can easily obtain significant suppression of $Q_{Ar}$ as well as  the upturn of the spectrum 
(see fig.~\ref{fig:rates}). There are, however,  restrictions which follow 
from other experimental results. 


{\it 1). Ar-production rate versus Ge-production rate.} 
A decrease of  $Q_{Ar}$ is accompanied by decrease of $Q_{Ge}$ (fig.~\ref{fig:rates}). 
Since the LMA prediction for $Q_{Ge}$ is close to  the
central experimental value a possible decrease of $Q_{Ge}$ is restricted. 
Let us consider this correlation in details.

The decrease of the $Ar$-production rate can be written as
\be
\Delta Q_{Ar}  = Q_{Ar}^{Be} \cdot \Delta P_{ee}^{Be} + 
Q_{Ar}^{int} \cdot \Delta P_{ee}^{int} + Q_{Ar}^{B} \cdot \Delta P_{ee}^{B},   
\label{ar}
\ee
where $Q_{Ar}^{Be} = 1.15$ SNU, $Q^{int}_{Ar} = 0.64$ SNU and 
$Q_{Ar}^{B} = 5.76$ SNU
are the contributions to the $Ar$-production rate from the
$Be$-flux, the fluxes of the intermediate energies ($pep$, $CNO$) 
and the $B$-neutrino flux according to 
SSM~\cite{ssm}.  Here $\Delta P_{ee}^{Be}$ is the  change of 
survival probability at $E_{Be}$,  $\Delta P_{ee}^{int}$ and $\Delta P_{ee}^{B}$ are the
changes of the effective (averaged over appropriate energy interval) 
survival probabilities for the intermediate energy 
fluxes and the boron neutrino flux  respectively.

The suppression of the $Ge$- production rate equals 
\be
\Delta Q_{Ge} \approx  Q_{Ge}^{Be} \cdot \Delta P_{ee}^{Be} +  
Q_{Ge}^{int} \cdot \Delta  P_{ee}^{int} +  Q_{Ge}^{B} \cdot \Delta P_{ee}^{B},  
\label{ga}
\ee
where $Q_{Ge}^{Be} = 34.2$ SNU, $Q^{int}_{Ge} = 11.7$ SNU and 
$Q_{Ge}^{B} = 12.1$ SNU 
are the contributions to the $Ge$-production rate  
for  the $Be$-neutrino flux,  the sum of $pep$- and
$CNO$- fluxes, and the $B$-neutrino flux correspondingly.  
$\Delta P_{ee}^{Be}$ is the same as in  (\ref{ar}),  whereas 
$\Delta P_{ee}^{int}$ and $\Delta P_{ee}^{B}$ are approximately equal to those  in (\ref{ar}).

The changes of rates are correlated:
\be
\Delta Q_{Ge} = A(R_{\Delta}, \alpha) \cdot \Delta Q_{Ar},  
\label{ar-ga1}
\ee
where $A$ is the constant which depends on the oscillation parameters.
If the  $Be$- ($\nu_e$) line is suppressed only, we would have $A^{Be} = \approx 30$.  
If the neutrino fluxes at the  intermediate
energies  are affected only, then $A^{int} \sim 18$,  for  the boron neutrino flux 
we find the smallest value $A^{B} \sim 2$.

In principle, the  decrease of the $Ge$-production rate can be compensated by increase
of the survival probability for the $pp$-neutrinos. This probability
is given approximately by the  average vacuum oscillations formula 
\be
P_{ee}(pp) \approx 1 - 0.5 \sin^2 2\theta.
\label{problo}
\ee
From Eq. (\ref{problo}) it follows that the increase of
$P_{ee}(pp)$ requires the decrease of mixing:
\be
\Delta \sin^2\theta  = - \frac{\Delta P_{ee}(pp)}{2\sqrt{2P_{ee}(pp) - 1}}.   
\label{xxss}
\ee
The SSM contribution of the $pp$-neutrinos to $Q_{Ge}$ equals
$Q_{Ge}^{pp} = 69.7$ SNU, therefore  to compensate $1\sigma$ ($ \sim 5$ SNU)
decrease of $Q_{Ge}$, one needs $\Delta P_{ee}(pp) = 0.07$. 
For this value of $\Delta P_{ee}(pp)$ eq.~(\ref{xxss}) gives  $\Delta \sin^2\theta = -
0.1$. However, a decrease of $\sin^2\theta$ is restricted by the high energy
data (SK, SNO). Indeed, the survival probability for the boron
neutrinos with $E > 5$ MeV is proportional to $\sin^2\theta$:
\be
P_{B} \approx a \sin^2\theta, ~~~~ a \approx 1.1,
\ee 
where the deviation of $a$ from 1 is due to  effects of the
upturn and the $\nu_e$ regeneration in the matter of the Earth.  So,
the survival probabilities for the $pp$- and $B$-neutrinos are
related:
\be
P_{pp} \approx 1 - \frac{2}{a} P_B + \frac{2}{a^2} P_B^2.
\label{lowhigh}
\ee 
For the best fit value of mixing ($\sin^2\theta \sim 0.285$) this
equality gives $\Delta P_{pp} \approx - 0.78 \Delta P_B$.  In turn,
the survival probability $P_B( > 5 ~{\rm MeV})$ is fixed by the CC/NC
ratio:
\be
\frac{CC}{NC} = \frac{P_{B}}{ 1 - \eta_s (1 - P_{B})}~, 
\ee
where $\eta_s$ is the sterile neutrino fraction in the state to which
$\nu_e$ transforms.  This relation does not depend on the original
Boron neutrino flux. The solar neutrino data restrict $\eta_s <
0.2$, and therefore the presence of sterile component allows us to reduce
the probability by a small amount only: $\Delta P_{ee} \sim (0.2 -
0.3) \eta_s < 0.06$.  Moreover, according to fig.~\ref{fig:psurv}, the
contribution of sterile neutrino to the high energy part of the
spectrum is even smaller than 0.2.

2).{\it ~ The  Ar-production rate versus the rates at SNO and SuperKamiokande.} 
For large $R_{\Delta}$ and $\sin \alpha$  the restriction appears  from 
the charged current (CC) - event rate at SNO and well as from the rate of events at SK 
and the spectra. 
For free boron neutrino flux  the suppression of $Q_{Ar}$ due to suppression 
of the boron electron neutrino flux  can be written as 
\be
\Delta Q_{Ar}  = 
Q_{Ar}^{B} f_B \Delta P_{ee}^{B, Ar},
\label{ar-B}
\ee
where $f_B \equiv F_B/F_B^{SSM}$ is the total boron neutrino flux in the units of the SSM flux. 
For the relative change of flux measured in CC-event, 
$\Delta [CC] \equiv \Delta F_{CC}/F_{CC}^{SSM}$ 
we have
\be
\Delta [CC]  =  f_B \Delta \tilde{P}_{ee}^{B},
\ee
where $\Delta \tilde{P}_{ee}^{B}$ is the  change of the effective survival  
probability for the SNO energy range. 
With decrease of $Q_{Ar}$ the rate $[CC]$ decreases; we find  
\be
\Delta [CC]  =  0.2 Q_{Ar}, 
\label{Ar-cc}
\ee
and this relation does not depend of $f_B$, so that for a given 
$Q_{Ar}$, the decrease of $[CC]$ can not be compensated by increase of $f_B$. 

Also the spectral information does not allow to strongly suppress $Q_{Ar}$. 


\section{Global Fit}

\begin{figure}[ht]
\centering\leavevmode
\epsfxsize=.8\hsize
\epsfbox{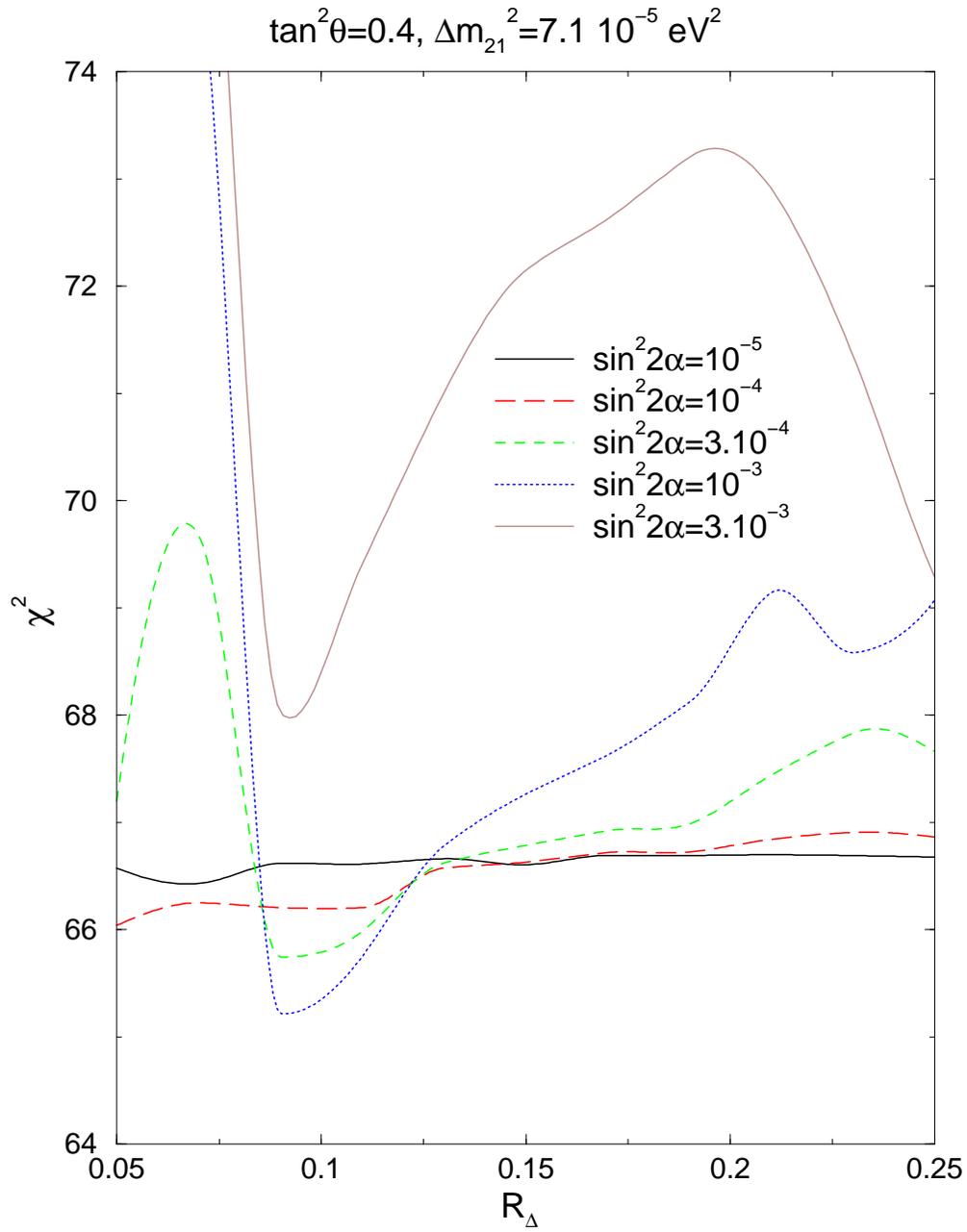}
\caption{The $\chi^2$ of the global fit of the solar neutrino data 
as a function of $R_\Delta$  for different values of the  
sterile-active mixing parameter $\sin^2 2\alpha$. 
We take  $\tan^2\theta = 0.4$ and 
$\Delta m^2_{12}=7.1\times 10^{-5}$ eV$^2$, 
}
\label{fig:chi2rdm}
\end{figure}

We have performed the global fit of the solar neutrinos data which takes into account 
the correlations of observables discussed in sec. 4.   
We use the same procedure of the fit as in our previous publications \cite{comb,ks}.
In fig.~\ref{fig:chi2rdm} we show the dependence of the $\chi^2$ on 
$R_{\Delta}$ for fixed value of $\Delta m_{21}^2$. 

The following comments are in order. 

1. According to  the fig.~\ref{fig:chi2rdm} the minimum $\chi^2_0 \sim 65.2$
is achieved for
\be
R_{\Delta} = 0.9 ~~~~\sin^2 2\alpha = 10^{-3}. 
\label{gfit}
\ee
It corresponds to the unmodified $Be$- flux but suppressed
$pep$- and $CNO$- fluxes. The upturn of the energy spectrum above
$5$ MeV is practically eliminated.
$\chi^2_0$ can be compared with $\chi^2 \sim 66.6$, for zero  $s$-mixing.
The improvements of the fit, $\Delta \chi^2 = 1.4$,   is not substantial.  
Notice however, that value $\chi^2_0$  is not the absolute minimum.
Furthermore,  one should not expect significant improvement of the fit
since the original pull was about $2\sigma$ only, and 
quality of the global fit is very good in both cases. 
Finally  with sterile neutrinos we have modified solution 
of solar neutrino problem with different set of  predictions for observables.

2. As follows from the  fig.~\ref{fig:chi2rdm} certain regions of  parameters
of the sterile neutrino are strongly disfavored or excluded already by existing data.
In particular, the region  $\sin^2 2\alpha = 3 \cdot 10^{-4}$ and
$R_{\Delta} < 0.07$ is excluded. It corresponds to  strong
suppression of the $Be$ electron neutrino flux.

In another strongly disfavored region:
$R_{\Delta} = 0.10 - 0.25$, $\sin^2 2\alpha > 10^{-3}$, 
one has substantial suppression of the $CC$-signal at SNO and SK as well as
distortion of the boron electron neutrino spectrum. For larger
values of sterile neutrino mass, $R_{\Delta} > 0.25$,
the dip shifts to higher energies and disappears. The 
conversion effects (and corresponding $\chi^2$) converge to the pure LMA solution case .

\section{Three scenarios}

\begin{figure}[ht]
\centering\leavevmode
\epsfxsize=.8\hsize
\epsfbox{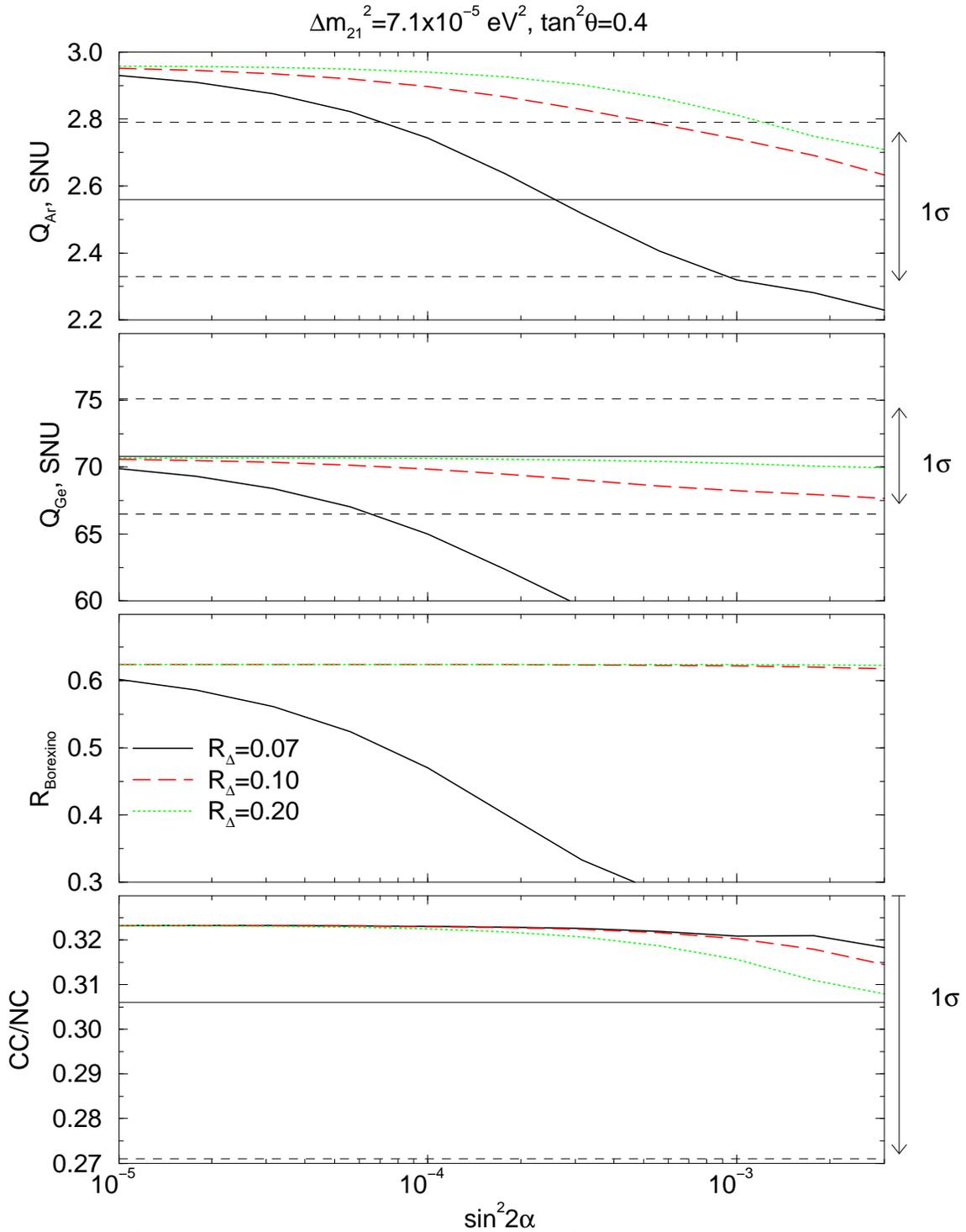}
\caption{
The $Ar$  production rate (upper panel), the  $Ge$ production rate
(second panel)
the suppression factor for the BOREXINO signal and the CC/NC ratio at SNO 
as  functions of $\sin^22\alpha$, for $\tan^2\theta = 0.4$ and   
$\Delta m^2_{21} = 7.1\times 10^{-5}$ eV$^2$.
}
\label{fig:rates}
\end{figure}

Three phenomenologically different scenarios can be realized depending
on the oscillation parameters,  and therefore on the  position and form 
of the dip. Three panels in the fig.~\ref{fig:rates}, which correspond
to different values of $R_{\Delta}$, illustrate these scenarios. 
Let us describe features of these three possibilities.

1). Narrow dip at low energies: the $Be$-line is in the dip. This corresponds
to $\sin^2 2\alpha < 10^{-4}$ and $R_{\Delta} < 0.08$ or 
\be
0.5 E_{Be} < E_s (n_c) < E_{Be},
\label{1case}
\ee
where $E_{Be} = 0.86$ MeV  is the energy of the $Be$-neutrinos (first panel in
fig.~\ref{fig:psurv} and solid line in fig.~\ref{fig:rates}).  The
lower bound (\ref{1case}) implies that the $pp$-neutrino flux is not
affected.  In this case the $Be$-line is  suppressed
most strongly; the $\nu_e$ fluxes of the intermediate energies ($pep$
and $CNO$ neutrinos) are suppressed weaker and the low energy part of
the boron neutrino spectrum measured by SK and SNO is practically
unaffected (see fig.~\ref{fig:rates}).

According to  fig.~\ref{fig:rates} the  value  of coefficient in Eq.~(\ref{ar-ga1}) 
$A = 24$. Taking the present $1\sigma$ errors, $0.23$ SNU and $5$ SNU, for the
Homestake and the combined Gallium result correspondingly, we find
that the central experimental value of $Q_{Ar}$ can be reached at the  
price of the $2\sigma$ decrease of  $Q_{Ge}$.

The best compromise solution would correspond to $\sin^2 2\alpha \sim
7 \cdot 10^{-5}$, when $Q_{Ar}$ is $1\sigma$ above the observation, and
$Q_{Ge}$ is $1\sigma$ below the observation. In this case the BOREXINO
rate reduces from 0.61 down to 0.48 of the SSM rate (see
sect.~\ref{sect:tests}).

For $E_s(n_c)$ being substantially smaller than $E_{Be}$,  the $Be$- line is
on the non-adiabatic edge of the dip and its suppression is weaker. In this case
larger values of  $\sin \alpha$ are allowed.

As we have discussed  in sec.4  variations of the LMA parameters and the original
boron neutrino flux do not allow us to compensate completely the changes
of the observables (which worsen the fit) in the case when
the $Be$-line is suppressed.

2). The dip at the intermediate energies:
\be
E_{Be} < E_s (n_c) < 1.4~ {\rm MeV}
\label{2case}
\ee
(see the second panel in fig.~\ref{fig:psurv} and the dashed lines in
fig.~\ref{fig:rates}).  The $Be$-line is out of the dip and therefore
unaffected. A decrease of $Q_{Ar}$ occurs due to suppression of the
$\nu_e$ components  of the $pep$- and $CNO$- neutrino fluxes.

In this case a decrease of $Q_{Ar}$ is accompanying by smaller
decrease of $Q_{Ge}$ in comparison with the previous case.  For small
enough mixing (so that the boron neutrinos are not affected strongly)
we get from fig.~\ref{fig:rates} $A = 15$ in the relation
(\ref{ar-ga1}).  For larger $\sin^2 2\alpha$ a  suppression of the boron
$\nu_{e}$ flux becomes substantial and $A$ decreases further: $A \sim
12$.  Now the value $Q_{Ar} = 2.8$ SNU, which is $1\sigma$ above the
observation, can be achieved by just $0.4\sigma$ reduction of
$Q_{Ge}$.

The BOREXINO signal due to the $Be$- flux is unchanged, 
and also the observable part of the boron neutrino flux is
affected very weakly.  Change of the CC/NC ratio is about 0.002.

The optimal fit (see fig.~\ref{fig:chi2rdm}) would correspond to 
$\sin^2 \alpha = 10^{-3}$, when $Q_{Ar}$ is diminished down to 2.75 SNU, 
at the same time $Q_{Ge} = 68$ SNU and CC/NC $= 3.22$ in agreement with the latest 
data \cite{salt}.

3). The dip at high energies: 
\be
E_s (n_c) > 1.6~ {\rm MeV} 
\label{2case2}
\ee
(see fig.~\ref{fig:psurv}, the  panel for $R_{\Delta} = 0.2$, and the dotted
lines in fig.~\ref{fig:rates}).  $Q_{Ar}$ is diminished due to
suppression of the low energy part of the boron neutrino spectrum. For
$\sin^2 \alpha = 10^{-3}$, we find $\Delta Q_{Ar} = 0.17$ SNU. At the
same time a decrease of the $Ge$-production rate is very small:
$\Delta Q_{Ge} \sim 0.5$ SNU which corresponds to  $A = (2 - 3)$
in eq. (\ref{ar-ga1}).  

At $\sin^2 \alpha = 10^{-3}$ there is already significant
modification of the observable part of the boron neutrino spectrum and
decrease of the total rate at SK and SNO.  
Also the CC/NC ratio decreases.  According to
fig.~\ref{fig:rates} at $\sin^2 2\alpha = 10^{-3}$, we have
$\Delta$(CC/NC) = 0.01.  Further increase of $R_{\Delta}$ will shift
the dip to higher energies, where the boron neutrino flux is
larger. This, however, will not lead to further decrease of $Q_{Ar}$
since the dip becomes smaller approaching the non-oscillatory region
(see fig.~\ref{fig:psurv}).  The BOREXINO signal ($Be$-line) is
unchanged.  So, the main signature of this scenario is a strong
suppression of the upturn and even a possibility to bend the spectrum
down.\\

Even for large $R_{\Delta}$ the influence of $\nu_s$   
on the KamLAND results is negligible due to very small mixing. 
In contrast to the solar neutrinos, for the  KamLAND experiment the matter effect 
on neutrino oscillations is very small and 
no enhancement of the $s$-mixing occurs. Therefore the  effect of $s$-mixing on  
oscillation probability is smaller than $\sin^2 2\alpha \sim 10^{-3}$. 
For this reason the KamLAND result has not been included  in the fit of data.  

\section{Further tests}
\label{sect:tests}

How one can check the described scenarios?

1) BOREXINO and KamLAND (solar) as well future low energy
experiments~\cite{LENS,MOON,XMASS,HERON,CLEAN,TPC,MUNU} can establish
the suppression of the $Be$-neutrino flux in comparison with the LMA
predictions, if the case 1) is realized.  
In BOREXINO and other experiments based on the $\nu e$-scattering 
the ratio of the numbers of
events  with and without conversion can be written as
\be
R_{Borexino} = P_{ee} (1 - r) + r - r P_{es},
\label{bore}
\ee
where $r \equiv \sigma (\nu_{\mu} e)/ \sigma(\nu_{e} e)$ is the ratio
of cross-sections.  Using
Eq. (\ref{delta}) we find an additional suppression of the BOREXINO rate in
comparison with the pure LMA case:
\be
\Delta R_{Borexino} \equiv  R_{Borexino}^{LMA} - R_{Borexino}  =  (1 - r)\Delta P_{ee}  + r P_{es} 
\approx  \Delta P_{ee} (1 + r \tan^2 \theta) .  
\label{dbor}
\ee
According to fig.~\ref{fig:rates},   
$R_{Borexino}^{LMA}$ can be diminished rather significantly. However, if the
prediction for $Q_{Ge}$ is  $2\sigma$  (or less) below the experimental
results, we find $R_{Borexino}^{LMA} > 0.4$ and $\Delta R_{Borexino} <
0.2$.  For the best fit value in the scenario 1):
\be 
R_{Borexino}^{LMA} \sim 0.5, ~~~(\Delta R_{Borexino} \sim 0.1).
\ee
Clearly, it will be difficult to establish such a difference.
Furthermore, an additional suppression is mainly due to conversion to the 
sterile neutrino and the problem is to distinguish the conversion
effect and lower original flux: the CC/NC ratio can not be used.  
Therefore not only high  statistics results
but also precise knowledge of the original fluxes is needed.  The
$pep$-flux is well known, however predictions of the $CNO$ neutrino
fluxes have larger uncertainties.

2). It may happen that the dip is at higher energies and the $Be$-
flux is unaffected.  In this case one expects significant suppression
of the $pep$- and $CNO$- fluxes.  Such a possibility can be checked
using combination of measurements from different experiments which are
sensitive to different parts of the solar neutrino spectrum.  The
radiochemical $Li$- experiment~\cite{lit1} has high sensitivity to the $pep$- and
$CNO$- neutrino fluxes
\cite{lit}. According to  SSM~\cite{ssm}, 
the $CNO$-neutrino contribution to the $Be$-production rate
in this 
experiment is $Q_{Be}^{CNO} = 14.2$ SNU of the
total rate $Q_{Be}^{CNO} = 52.3$ SNU,  so that $Q_{Be}^{CNO}/Q_{Be} =
0.27$. For the $Cl$- and $Ga$- experiments the corresponding ratios
are substantially smaller: $Q_{Ar}^{CNO}/Q_{Ar} = 0.05$ and $Q_{Ge}^{CNO}/Q_{Ge} = 0.07$.
For the $pep$-neutrinos we get $Q_{Be}^{pep}/Q_{Be} = 0.176$.

Precise measurements of $Q_{Be}$ and $Q_{Ge}$ and independent
measurements of the $B$, $pp$  and  $Be$ neutrino fluxes and subtraction of
their contributions from $Q_{Be}$ and $Q_{Ge}$ will allow to determine
the $CNO$- electron neutrino fluxes.  In general, to measure
oscillation parameters and to determine the original solar neutrino
fluxes one will need to perform a combined analysis of results from $Ga$-,
$Cl$-, $Li$- experiments as well as  the dedicated low energy
experiments~\cite{LENS,MOON,XMASS,HERON,CLEAN,TPC,MUNU}.  Of course,
new high statistics $Cl$-experiment would clarify the  situation
directly.

\begin{figure}[ht]
\centering\leavevmode
\epsfxsize=.8\hsize
\epsfbox{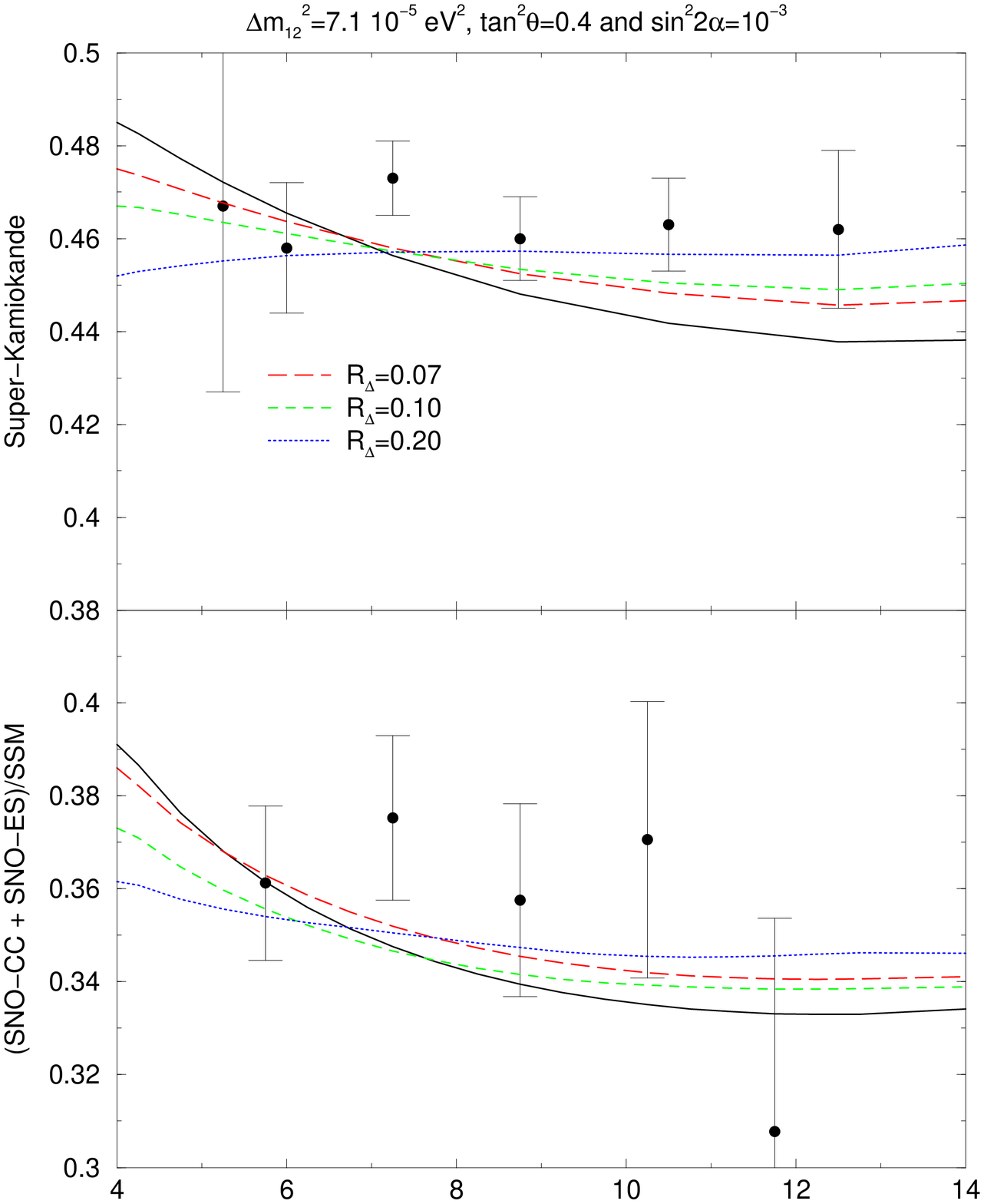}
\caption{The spectrum distortion ($N^{osc}/N^{SSM}$) 
at Super-Kamiokande (upper panel)
and SNO (lower panel) for different values of the mass ratio $R_{\Delta}$ 
and for the sterile-active mixing 
$\sin^2 2\alpha = 10^{-3}$. The solid lines correspond to 
the pure LMA case (no sterile neutrino).   
Normalization of spectra have been chosen
to minimize $\chi^2$ fit of spectrum for each case. We show also the
Super-Kamiokande and SNO experimental data points with statistical errors 
only.}
\label{fig:skspec}
\end{figure}

3). For $R_{\Delta} \sim 0.1 - 0.2$ and $\sin^2 2\alpha \sim 10^{-3}$ 
one expects significant suppression of the low
energy part of the $B$- neutrino spectrum.  As follows from 
figs.~\ref{fig:skspec}, at 5 MeV an  additional 
suppression due to sterile neutrino can reach (10 - 15)\% both in SK and SNO. 
The spectra with the  $s$-mixing give slightly better fit to the data.
Notice that there is no turn down of the SNO spectrum
for $R_{\Delta} = 0.2$ and $\sin^2 \theta_{13} = 10^{-3}$
due to an additional contribution from the $\nu -  e$ scattering.
Precision measurements of  shape of the spectrum in
the low energy part ($E < 6 - 8$ MeV) will give crucial checks
of the described possibility.

4). In supernovae, neutrinos are produced at densities far above the
LMA resonance density and  propagation is adiabatic in the LMA
resonance.  So, even for very small 1-3 mixing ($\sin^2 \theta_{13} > 10^{-4}$)  
the adiabatic conversion $\nu_e \rightarrow \nu_2$ is realized without any effect of
sterile neutrino (as in the case described in Eq. (\ref{esmall})).
If, however, the sterile level crosses the second level $\lambda_2^{LMA}$
one may expect some manifestations of the sterile neutrino in the
$\nu_e$ channel, provided that the mass hierarchy is inverted or the
1-3 mixing is very small ($\sin^2 \theta_{13} < 10^{-4}$).

5). Smallness of mixing of the sterile neutrino allows one to satisfy  the
nucleosynthesis bound: such a neutrino does not equilibrate in the
Early Universe.  For this reason sterile neutrinos also do not
influence the large scale structures formation in the Universe.

6). A very small $s$-mixing means that the width of  $s$-resonance
is also very small.  In the density scale $\Delta n/n = \tan 2\alpha
\sim 10^{-2}$. Therefore 1\% density perturbations can strongly affect 
conversion in the $s$-resonance~\cite{pert}. If density perturbations
(or density profile) change in time, this will induce time variations
of neutrino signals. Since the effect of $s$-resonance is small, one
may expect $~10\%$ (at most) variations of the $Ga$- and $Ar$-
production rates.

It seems that further precision measurements of the solar neutrino
signals are the only possibility to check the suggested scenarios.\\

\section{Conclusions}

1. The low (with respect to the LMA prediction) value of the
$Ar$-production rate measured in the Homestake experiment and/or
suppressed upturn of the spectrum at low energies in SK and SNO can be
explained by introduction of the sterile neutrino which mixes very
weakly with the active neutrinos.

\noindent
2. The mixing of  sterile neutrino leads to appearance of the dip in
the survival probability in the interval of intermediate energies $E =
0.5 - 5$ MeV.  The survival probability in the non-oscillatory and
vacuum ranges is not modified (if sterile level crosses $\lambda_1^{LMA}$).

\noindent
3. Depending on value of $R_{\Delta}$, that is,
on a  position of the dip, three phenomenologically different
scenarios are possible:

(i) the $Be$-neutrino line in the dip; 

(ii) strong suppression of the $pep$- and $CNO$- neutrino fluxes, and the 
$Be$-neutrino line out of the dip;

(iii) suppression of the boron flux only.

The best global fit of the solar neutrino data corresponds to the 
unsuppressed $Be$-line, but strongly suppressed $pep$- and $CNO$-
neutrino fluxes. Such a scenario  requires $\sin^2 2\alpha \sim 10^{-3}$ and
$R_{\Delta} \sim 0.1$.  It predicts also an  observable suppression
of the upturn of the spectrum at SK and SNO.

\noindent
4.  The present experimental results as well as relations between
observables  restrict substantially possible effects of the dip induced by the
$s$-mixing.

\noindent
5. The presence of $s$-mixing can be established by  future precise
measurements of the $Be$-, $pep$-, $CNO$- neutrino fluxes in BOREXINO
\cite{bor} and KamLAND, as well as by measurements of  the low energy part of the Boron neutrino
spectrum ( $< 5 - 6$ MeV) in SNO.  Study of the solar
neutrinos seems to be  the only possible way to test the scenarios described in
this paper. There is no observable effects in laboratory experiment,
as well as in astrophysics and cosmology.

\noindent
6. We have performed a general study of the effect of mixing with
additional neutrino states (see the Appendix). Only in the case when the sterile 
neutrino  level crosses both the  LMA levels, the effect of additional mixing can enhance
the survival probability.  This case is not realized, however, for
additional sterile neutrino.  In all other cases an additional mixing
leads to suppression of survival probability.

\noindent
7. Even precise measurements of the high energy part of the solar
neutrino spectrum may not be enough to reconstruct the energy profile
of the effect at low energies.  So, the low energy solar experiments
are needed and they may lead to important discoveries.

\section{Note added}

Since the time we posted our paper on hep-ph, some new publications 
have appeared which are relevant for this study. 

1). Lower value of  the cross-section $^{14}N (p, \gamma)^{15}O $ 
measured by the LUNA experiment~\cite{LUNAex} leads to decrease 
of the predictions for the $Ar$-production rate are 
by  $\Delta Q_{Ar} = - 0.1$ SNU \cite{LUNAth} (see our footnote 1). This 
reduces  a difference of the LMA prediction and the Homestake result by 
about $0.5\sigma$. Notice that at the same time 
the $Ge$-production rate is dimished by $\Delta Q_{Ge} = 2$ SNU. 

2). Larger values of the  $^7 Be (p,\gamma) ^8B$ cross-section 
obtained in the recent measurements lead to 
significant increase of the predicted boron neutrino flux. 
Now the predicted flux is larger than than extracted from  
the NC event rate measured at SNO: $f_B = 0.88 \pm 0.04 (exp) \pm 
0.23 (theor)$  
\cite{BP04}. Being confirmed this may testify for partial conversion 
of the produced $\nu_e$ to 
sterile neutrino thus supporting scenario suggested in this paper.

\section*{Acknowledgments}

One of us (PCH) would like to thank FAPESP for financial support.

\section*{Appendix: Profile of the effect and new neutrino states}

As we have established in the previous sections,   mixing with
sterile neutrino can modify the LMA energy profile, namely,  suppress
the survival probability  in certain energy range.  Here we present a
general consideration of possible modifications of the LMA energy profile by 
mixing with new neutrino states.  

In general, an  introduction of new states leads to decrease of
$P_{ee}$, since new channels  open for disappearance of $\nu_e$.
What are conditions for increase of $P_{ee}$?

In the LMA case  the survival probability at low energies and at
high energies are uniquely related (\ref{lowhigh}).  So, 
in principle,  measurements at high energies ($E > 5$ MeV) allow 
to reconstruct the profile at low energies provided that 
$\Delta m^2$ is well determined. The latter can be achieved by KamLAND.    
Mixing with new states can change this high - low energy relation.

In assumption that  the coherence of  all  mass eigenstates is lost on
the way to the Earth we can write the $\nu_e$-survival probability
after propagation in the Sun as
\be   
P_{ee} = \sum_i a_i |U_{ei}|^2~,
\label{gen1}
\ee
where $a_i \equiv  |\langle \nu_i | \nu_f \rangle|^2 $ is the probability to find the mass state 
$i$ in the final
state and $U_{ei} \equiv \langle \nu_e| \nu_i \rangle$ is the mixing
parameter.  The quantities in eq. (\ref{gen1}) satisfy the
normalization conditions:
\be   
\sum_i  a_i = 1,~~~~~ \sum_i |U_{ei}|^2 = 1 .           
\ee

At low energies, where neutrino conversion is due to the vacuum
oscillations, the admixtures of mass eigenstates are not changed and
flavors are determined.  Contributions from two different mass eigenstates add
incoherently.  In the $2\nu$- case we have $a_1 = \cos^2 \theta$, $a_2
= \sin^2 \theta$, $U_{e1} = \cos \theta$, $U_{e2} = \sin \theta$, and
consequently, $P_{ee} = \cos^4 \theta + \sin^4 \theta$.  

The only way to increase  $P_{ee}$  in vacuum, 
would be to restore the coherence (at least partially) of the two
contributions, or decrease the mixing.  In general, one should
concentrate the electron flavor on one of the mass eigenstates and
increase its admixture.

Suppose  additional neutrino states also produce the vacuum
oscillation effect (no level crossing).  Mixing of these new states
with $\nu_e$ leads to  decrease of $|U_{e1}|^2$
or/and $|U_{e2}|^2$, as well as $a_1$ and $a_2$, and one can easily
show that
\be
P_{ee}(2) \geq P_{ee} (2 + n).
\ee
So, new states can lead to decrease of $P_{ee}$ only.
 
Matter effects change $a_i$. At high energies for the $2\nu$ case we
get $a_1 \approx 0$ and $a_2 \approx 1$.  Let $|U_{ei}|_{min}$ and
$|U_{ei}|_{max}$ be the largest and smallest mixing parameters
correspondingly.  Then it is easy to prove inequality
\be
|U_{ei}|_{min}^2 \leq P_{ee} \leq |U_{ei}|_{max}^2.
\label{ineq}
\ee
So, the only way to increase $P_{ee}$ is to change the admixtures of
the mass states in such a way that $a_i$, which corresponds to the
largest $|U_{ei}|$, increases.

Let us consider one additional neutrino level (state) which mixes
weakly with the LMA levels.  Due to small mixing the LMA levels do not
change significantly.  If new (predominantly sterile) level crosses
one of the LMA level only and $P(i)$ is the probability that neutrino state does not 
transit to new level in this crossing, then
\be   
P_{ee} \approx P(1) a_1 |U_{e1}|^2 + a_2 |U_{e2}|^2~.
\ee
Here we put $|U_{e0}| \approx 0$.  Since $P(1) < 1$, the probability
decreases as we have found in the specific case discussed in this
paper.  Similarly, if the new level crosses the second LMA level, 
the survival probability
\be   
P_{ee} \approx a_1 |U_{e1}|^2 + P(2) a_2 |U_{e2}|^2~, 
\label{gen2}
\ee
decreases. 
Notice that if neutrino is produced far above the LMA resonance, so
that $a_1 \approx 0$, the probability equals $P_{ee} \approx P(2) a_2
|U_{e2}|^2$ and for small $P(2)$ (adiabaticity) 
the probability  $P_{ee}$ can be strongly suppressed.

To enhance $P_{ee}$ the new level should cross both LMA levels (in
this case formulas above are not valid).  If both crossings are
adiabatic, the following transitions occur:
\be 
\nu_e \approx \nu_{2m} \rightarrow \nu_{0m} \rightarrow \nu_{1m}~. 
\ee
So that  $P_{ee} = \cos^2 \theta$. If transitions are partially
adiabatic, we find $\sin^2 \theta < P_{ee} < \cos^2 \theta$.  
Thus,  the admixture of the mass state with the largest fraction  of the
electron neutrinos is enhanced.  However, to get such a double
crossing, the new level should have stronger dependence on the density 
than the dependence of the electron neutrino
level. That is, the corresponding matter potential should be large:
$V_x > V_e$.  This is excluded:  an additional sterile
neutrino level can cross only one LMA level, thus leading to
suppression of the survival probability.  The results obtained in this
paper are robust.

\newpage

\newpage

\newpage

%
%

\newpage



\begin{thebibliography}{99}

\bibitem{kam} K. Eguchi et al, (KamLAND collaboration),
Phys. Rev. Lett, {\bf 90} 021802 (2003). 

\bibitem{salt} SNO collaboration (Q. R. Ahmad {\it et al.}),
nucl-ex/0309004.


\bibitem{w1} L. Wolfenstein, Phys. Rev. D{\bf 17} (1978) 2369; 
in {\it ``Neutrino -78"}, Purdue Univ. C3, 1978.
  
\bibitem{ms} S. P. Mikheyev and A. Yu. Smirnov, Sov. J. Nucl. Phys.
{\bf 42} (1985) 913;  Nuovo Cim. {\bf C9} 17, (1986);
Zh. Eksp. Teor. Fiz. {\bf 91}, 7 (1986),
[Sov. Phys. JETP {\bf 64}, 4  (1986)].

\bibitem{lma} J. N. Bahcall,  P. I. Krastev, A. Yu. Smirnov,
Phys. Rev. D {\bf 60} 093001 (1999);
M.C. Gonzalez-Garcia, P.C. de Holanda, Carlos Pena-Garay,
J.W.F. Valle, Nucl. Phys. B {\bf 573},  3 (2000).

\bibitem{solar} A. B. Balantekin and H. Y\"uksel, hep-ph/0309079; 
G.L. Fogli, E. Lisi, A. Marrone, A. Palazzo, hep-ph/0309100; 
M. Maltoni, T. Schwetz, M. A. Tortola, J.W.F. Valle,
hep-ph/0309130 (v.2); P. Aliani, V. Antonelli, M. Picariello,
E. Torrente-Lujan, hep-ph/0309156; 
P. Creminelli, G. Signorelli, A. Strumia, hep-ph/0102234, v.5,
Sept. 15  (2003);
A. Bandyopadhyay, S. Choubey,  S. Goswami, S. T. Petcov,  D.P. Roy,
hep-ph/0309174.

\bibitem{comb}P. C. de Holanda and A. Yu. Smirnov, 
hep-ph/0309299.

\bibitem{hom} B. T. Cleveland, et al., Astrophys.J. {\bf 496}, 505 (1998).  

\bibitem{ssm}J. N. Bahcall, M. H. Pinsonneault and S. Basu,
Astrophys. J., {\bf 555} (2001) 990.

\bibitem{ks} P.I. Krastev, A.Yu. Smirnov, Phys. Rev. {\bf D65},  
073022 (2002).  

\bibitem{sk} Super-Kamiokande Collaboration (S. Fukuda et al.),
Phys. Rev. Lett. {\bf 86} (2001), {\it ibidem}, {\bf 86},  5656 (2001).  

\bibitem{sno} SNO Collaboration (Q.R. Ahmad et al.), Phys. Rev. Lett. 
{\bf 89}, 011301 (2002), Phys. Rev. Lett. {\bf 89}, 011302, (2002).

\bibitem{vari} P.A. Sturrock, G. Walther and M. S. Wheatland,
Astrophys. J. {\bf 491} , 409 (1997);
{\it ibidem} {\bf 507}, 978 (1998); P.A. Sturrock and M. A. Weber, 
Astrophys. {\bf J. 565}, 1366 (2002).
 P.A. Sturrock, astro-ph/0304148; hep-ph/0304106.
%
P. A. Sturrock and M. A. Weber, Astrophys. J. {\bf 565}, 1366 (2002);
%
A. Milsztajn, hep-ph/0301252;
%
P. A. Sturrock, hep-ph/0304073; hep-ph/0304106.
%
D. O. Caldwell and P. A. Sturrock, hep-ph/0305303.

\bibitem{three} Incomplete list: 
T.K. Kuo, James Pantaleone, Phys. Rev. Lett. {\bf 57}, 1805 (1986), 
Phys. Rev. {\bf D35} 3432 (1987),   
A. Yu. Smirnov, in proc. of the Int. Symposium on Neutrino Astrophysics 
``Frontiers of Neutrino Astrophysics'', October 19-22 (1992) 
Takayama, Japan, Eds. Y. Suzuki and K. Nakamura, p.105.,  
Q.Y. Liu, S.T. Petcov, Phys. Rev. {\bf D56}, 7392 (1997),  
K.S. Babu, Q.Y. Liu, A. Yu. Smirnov,  Phys. Rev. {\bf D57}, 5825 (1998).    

\bibitem{3st}
V. Berezinsky, M. Narayan, F. Vissani,  Nucl. Phys. {\bf B658}, 254 (2003). 
P.C. de Holanda,   A.Yu. Smirnov,  hep-ph/0211264.  

\bibitem{sage} SAGE Collaboration (J.N. Abdurashitov et al.),  
J.Exp.Theor.Phys. {\bf 95},   181 (2002), [Zh.Eksp.Teor.Fiz. {\bf
122},  211 (2002)] 
e-Print Archive: astro-ph/0204245.  

\bibitem{gno} GNO Collaboration (T. A. Kirsten), Nucl. Phys. B 
(Proc. Suppl.) {\bf 118} (2003) 33. 


\bibitem{bor} BOREXINO collaboration, G. Alimonti, {\it et al.}, 
Astropart. Phys. {\bf 16} 073022 (2002). 

\bibitem{LENS} R.S. Raghavan,  Phys. Rev. Lett. {\bf 78}, 3618 (1997).  
 
\bibitem{MOON} H. Ejiri et al., Phys.Rev. Lett. {\bf 85}, 2917 (2000).

\bibitem{XMASS} Y. Suzuki, Low Nu2 workshop, Tokyo 2000,\\ 
http://www-sk.icrr.u-tokyo.ac.jp/neutlowe/.

\bibitem{HERON} B. Lanou, in  Low Nu3 Heidelberg, Germany, 
http://www.mpi-hd.mpg.de/nubis/.

\bibitem{CLEAN} D. N. McKinsey and J. M. Doyle, J. of Low Temp. Phys. 
{\bf 118} 153 (2000). 

\bibitem{TPC} G Bonvicini et al., hep-ex/0109199, hep-ex/0109032. 

\bibitem{MUNU}C. Broggini, Low Nu3 Heidelberg, Germany, 
http://www.mpi-hd.mpg.de/nubis/.  

\bibitem{lit1} J. N. Bahcall,  Phys. Lett., {\bf 13},  332 (1964)
V. A. Kuzmin and G. T. Zatsepin, Proc. Int. Conf on Cosmic Rays,
Jaipur, India, 1965 paper Mu-Nu 36, p. 1023.

\bibitem{lit} See for recent discussion:  
A. Kopylov, V. Petukhov, hep-ph/0301016, hep-ph/0306148. 

\bibitem{pert} P.I. Krastev, A.Yu. Smirnov, Mod. Phys. Lett. 
{\bf A6}, 1001 (1991).

\bibitem{LUNAex} LUNA Collaboration (A. Formicola {\it et al}.), 
nucl-ex/0312015. 

\bibitem{LUNAth} S. Degl'Innocenti, G. Fiorentini, 
B. Ricci, F.L. Villante, astro-ph/0312559. 

\bibitem{BP04} J. N. Bahcall and M. H. Pinsonneault, 
astro-ph/0402114. 


\end{thebibliography}
\end{document}